\documentclass{svjour3}
\smartqed
\usepackage{amsmath,graphicx}

\journalname{Journal of Low Temperature Physics - QFS2009}

\begin{document}

\title{
Determination of the mosaic angle distribution of Grafoil
platelets using continuous-wave NMR spectra
}

\author{
S. Takayoshi$^1$ \and Hiroshi Fukuyama$^2$
}

\institute{
1:Department of Physics, Graduate School of Science,
The University of Tokyo, 7-3-1 Hongo, Bunkyo-ku, Tokyo 113-0033, Japan\quad
Tel., Fax:+81 3 5841 8364\\
\emph{Present address:Institute for Solid State Physics, The University of Tokyo,
5-1-5 Kashiwanoha, Kashiwa, Chiba 277-8581, Japan}\quad
Tel.:+81 4 7136 3277\\
\email{shintaro@issp.u-tokyo.ac.jp}\\
2:Department of Physics, Graduate School of Science,
The University of Tokyo, 7-3-1 Hongo, Bunkyo-ku, Tokyo 113-0033, Japan\\
\email{hiroshi@phys.s.u-tokyo.ac.jp}
}

\date{Received: date / Accepted: date}

\maketitle

\begin{abstract}
We described details of a method to estimate
with good accuracy the mosaic angle distributions of 
microcrystallites (platelets) in exfoliated graphite like Grafoil which is commonly used
as an adsorption substrate for helium thin films.
The method is based on analysis of resonance field shifts in continuous-wave (CW)
NMR spectra of $^{3}$He ferromagnetic monolayers making use of the large nuclear 
polarization of the adsorbate itself.
The mosaic angle distribution of a Grafoil substrate analyzed in this way can
be well fitted to a gaussian form with a $27.5\pm2.5$ deg spread.
This distribution is quite different from the previous estimation based on 
neutron scattering data which showed an unrealistically large isotropic powder-like 
component.
\keywords{Two dimensional helium-3 \and NMR \and Substrate effects}
\PACS{67.30.-n \and 67.30.ej \and 67.30.er}
\end{abstract}

\section{Introduction}
\label{sec:intro}
Atomically thin $^{3}$He film adsorbed on graphite is an ideal model system for 
studying strong correlation effects in two-dimensional (2D) fermions.
In this system, the correlation can be tuned in a wide range by changing $^{3}$He 
areal density ($\rho$).
Intensive studies have been done especially in the density region around
the 4/7 commensurate phase in the second layer \cite{Fukuyama_JPSJ}.
Grafoil, an exfoliated graphite, is an adsorption substrate commonly used
in these kinds of experiments because of its large surface area ($\approx$ 20 m$^2$/g)
and moderate thermal conductivity. However, Grafoil consists of
a lot of microcrystallites (platelets) with mosaic angle distributions.
Thus, the experimental results sometimes suffer from the heterogeneity
effects of substrate.
For example, the non-exponential decay of NMR spin-echo signals
recently observed near the 4/7 phase \cite{Takayoshi_JPCS}
and the linear Larmor frequency-dependence of transverse spin relaxation rate
($T_{2}^{-1}$) \cite{Cowan_JJAP} are believed to be due to such heterogeneity effects.
Therefore, it is important to obtain distributions of the platelet size
and the mosaic angle spread accurately enough for each batch if possible.
So far, there exist only a few experimental determinations of the platelet distributions, 
and it is desired to cross-check the results by different methods.

In this paper, we demonstrate that the mosaic angle distribution can be determined
with good accuracy by analyzing continuous-wave (CW) NMR spectra and
consequently that the distribution claimed
by the previous neutron scattering experiments \cite{Taub_PRB} is not correct.

\section{Previous studies}
\label{sec:prev}

\begin{figure}[b]
\begin{minipage}{0.48\hsize}
\begin{center}
\includegraphics[width=0.9\linewidth]{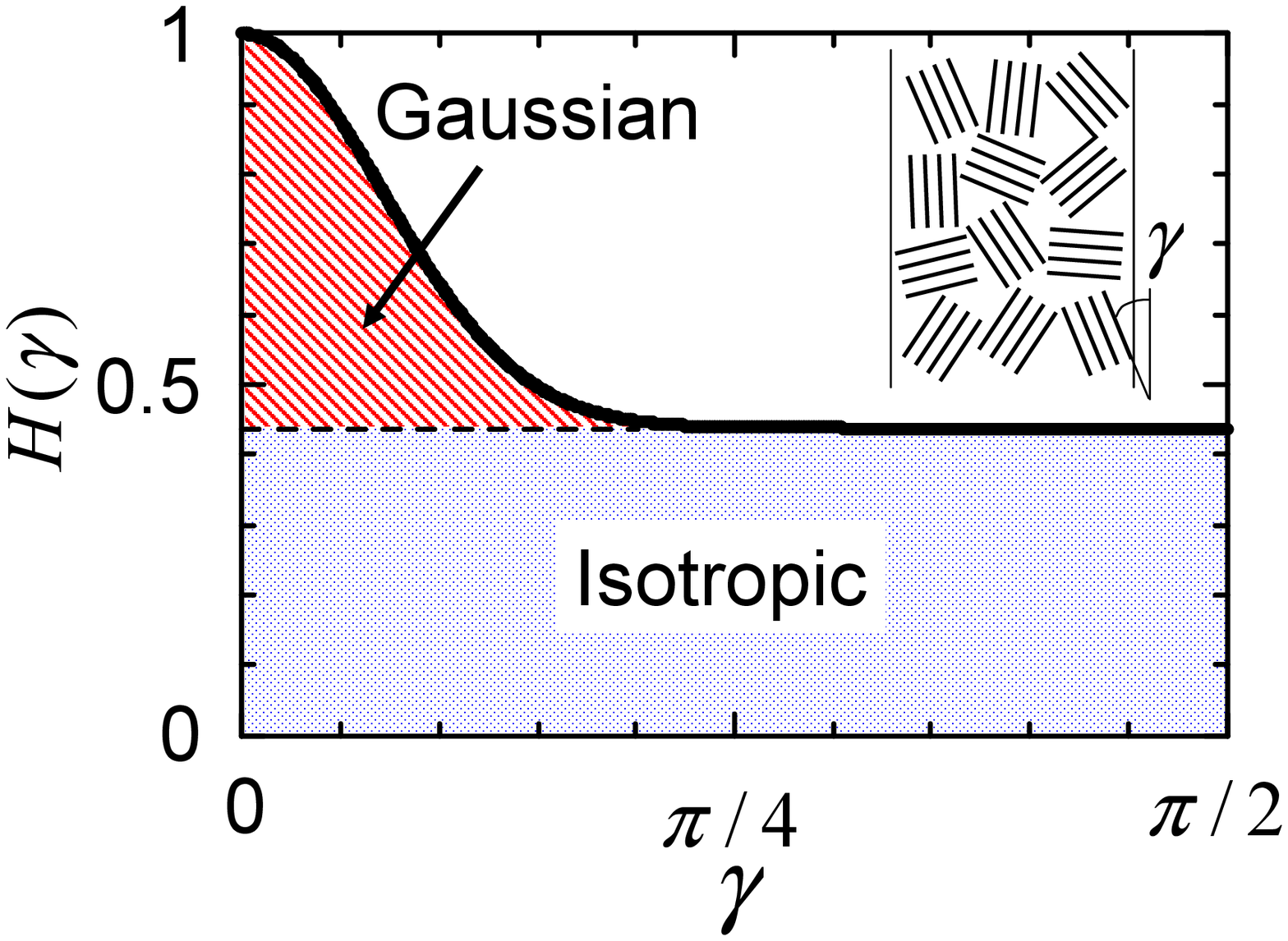}
\end{center}
\caption{(Color online) The orientational distribution of platelets in Grafoil
derived from the neutron diffraction experiments \cite{Taub_PRB}.}
\label{fig:distribution}
\end{minipage}
\hspace*{0.02\hsize}
\begin{minipage}{0.48\hsize}
\begin{center}
\includegraphics[width=0.9\linewidth]{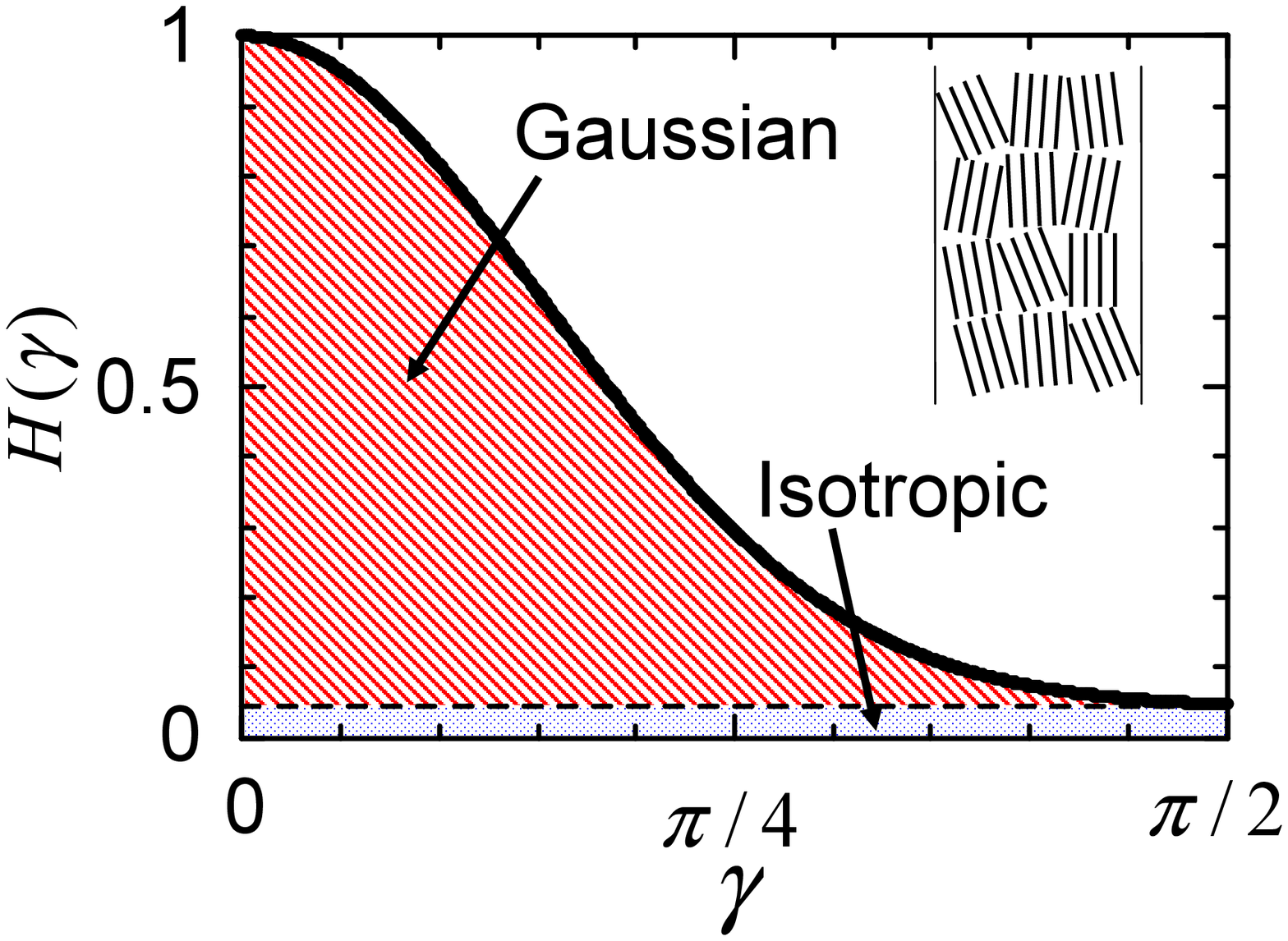}
\end{center}
\caption{(Color online) The orientational distribution derived from
analyzing CW-NMR spectra by our new method.}
\label{fig:distribution_new}
\end{minipage}
\end{figure}

The distribution of the platelet size in Grafoil estimated from scanning tunneling 
microscopy (STM) measurements \cite{Niimi_PRB} is widely spread below 100 nm.
The mosaic angle distribution of platelets was determined by the
neutron scattering experiments \cite{Taub_PRB} as follows.
For 2D crystals, the intensity diffracted at an angle $2\Theta$
by the Bragg reflection with the plane $(h,k)$ is represented as
\begin{equation}
I_{hk}=N\frac{m_{hk}|F_{hk}|^{2}f(\Theta)^{2}\mathrm{e}^{-2W}}{(\sin\Theta)^{3/2}}
\left(\frac{L}{\pi^{1/2}\lambda}\right)^{1/2}\mathcal{F}(a).
\end{equation}
Here, $N,m_{hk}$ and $f(\Theta)$ are a normalization constant,
the multiplicity of the $(hk)$ reflection and the molecular form factor, respectively.
$\mathrm{e}^{-W}$ is the Debye-Waller factor, and
\begin{equation}
\mathcal{F}(a)\equiv\int_{0}^{\infty}
\mathrm{e}^{-(x^{2}-a)^{2}}\mathrm{d}x,
\end{equation}
where $a=(2\pi^{1/2}L/\lambda)(\sin\Theta -\sin\Theta_{hk})$
and $\Theta_{hk}=\sin^{-1}\lambda/2d_{hk}$.
$\lambda, d_{hk}$ and $L$ are the wavelength, 2D plane spacing for the
$hk$-reflection and a parameter defining the average size of the 
diffracting arrays, respectively.
Using the orientational distribution function of Grafoil platelets
$H(\gamma(\Theta))$, the instrumental resolution factor $R(\Theta)$ and $I_{hk}$,
the angle dependence of the scattering intensity $g_{hk}$ is expressed as

\begin{equation}
g_{hk}=\int R(\Theta-\Theta ')H(\gamma(\Theta '))I_{hk}(\Theta ')
\mathrm{d}\Theta '.\label{eq:nscat_int}
\end{equation}
Thus $H(\gamma)$ can be determined by comparing measured diffraction peaks
with ones calculated through eq.(\ref{eq:nscat_int}).
Taub \textit{et al.} \cite{Taub_PRB} assumed $H(\gamma)$ as
the sum of gaussian and some constant 

\begin{figure}[b]
\begin{center}
\includegraphics[width=0.94\linewidth]{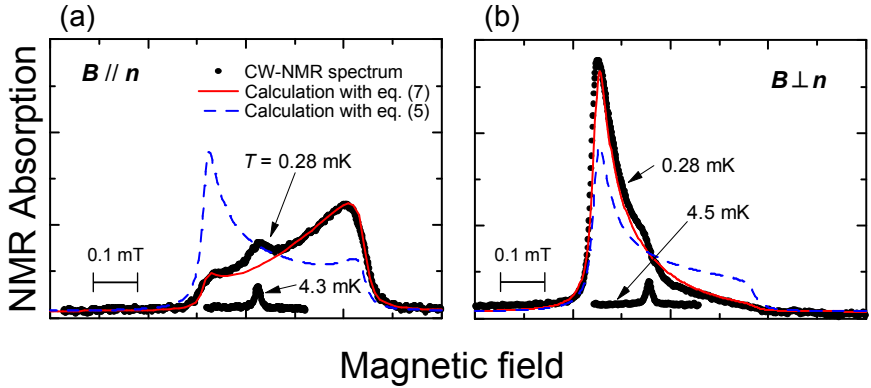}
\end{center}
\caption{(Color online) CW-NMR absorption spectra of the second-layer $^{3}$He adsorbed on 
Grafoil for two different field directions: (a) $\boldsymbol{B}\parallel\boldsymbol{n}$
and (b) $\boldsymbol{B}\perp\boldsymbol{n}$. 
The points are experimental data for the $\rho =$ 21.5 nm$^{-2}$ sample 
at temperatures indicated (from Ref. \cite{Schiffer_JLTP}).
The solid and dashed lines are calculated spectra with eq. (\ref{eq:dist_nmr}) and 
eq. (\ref{eq:dist_neu}), respectively.}
\label{fig:cw}
\end{figure}

\begin{equation}
H(\gamma)=H_{0}+H_{1}\exp
\left(-\frac{\gamma^{2}}{2\delta^{2}}\right),
\label{eq:dist_fn}
\end{equation}
and obtained
\begin{equation}
H_{0}/H_{1}=0.78,   \delta =12.7 \;\mathrm{deg}
\label{eq:dist_neu}
\end{equation}
from their own neutron diffraction data for Ar monolayer adsorbed on Grafoil.
As is shown in Fig. \ref{fig:distribution}, this has a large isotropic powder-like 
component $H_{0}$ in addition to the angle dependent term with $H_{1}$.
However, this is hard to believe since it is too isotropic judging from
the existing ample experimental properties of thin films adsorbed on Grafoil and 
of the substrate itself, e.g., the thermal conductivity \cite{Dillon_JLTP}, that all
show the significant anisotropy.

\section{New analysis of the previous CW-NMR data}
\label{sec:anal}

In this section, we introduce a quite different method to determine the mosaic angle 
distribution of Grafoil platelets based on an analysis of CW-NMR spectra of $^{3}$He 
thin films adsorbed on this substrate.
Here we use the NMR data for the high density second-layer $^{3}$He 
at very low temperatures, where the nuclear spins are ferromagnetically aligned
along the direction of applied magnetic field $\boldsymbol{B}$, measured 
by Schiffer \textit{et al.} 
\cite{Schiffer_JLTP}.
The NMR frequency shift due to the demagnetization effect is expressed as 
$\Delta\nu =\Delta\nu_{\mathrm{max}} P(1-3\cos^{2}\alpha)$, where 
$\Delta\nu_{\mathrm{max}}$ is the maximum positive value of frequency shift,
$P$ is the degree of spin polarization
and $\alpha$ is the angle between $\boldsymbol{B}$ and 
the vector normal to the average plane of Grafoil $\boldsymbol{n}$ \cite{Bozler_PRB}.
Since the inclination of each platelet about the average plane
has two degrees of freedom, it is designated by two angular variables,
$\theta$ and $\phi$, in the spherical coordinate with
$\boldsymbol{n}$ being the positive direction of $z$-axis.
Then, the NMR line shape about the field corresponding to the Larmor frequency
$B_{0}$ is represented as

\begin{figure}[b]
\begin{center}
\includegraphics[width=0.94\linewidth]{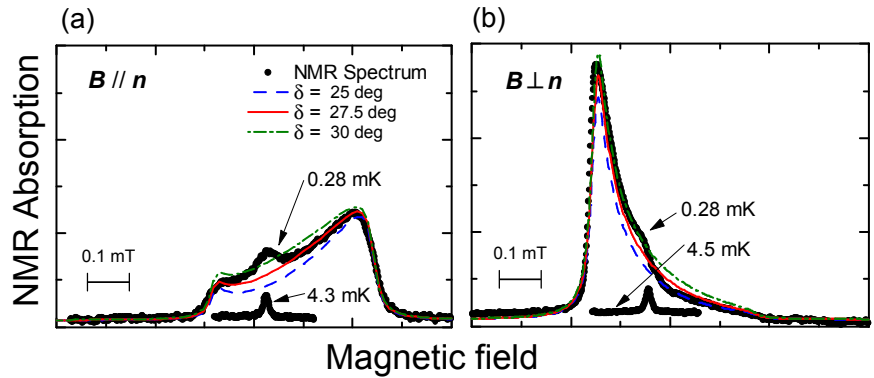}
\end{center}
\caption{(Color online) CW-NMR spectra of the second-layer $^{3}$He adsorbed on Grafoil. 
The dashed, solid and dash-dotted lines are calculated spectra with
$\delta = 25, 27.5, 30$ deg, respectively.
$H_{0}/H_{1}$ is fixed at 0.045. 
Otherwise, the same as Fig. \ref{fig:cw}.}
\label{fig:cw_param_delta}
\end{figure}

\begin{figure}[b]
\begin{center}
\includegraphics[width=0.94\linewidth]{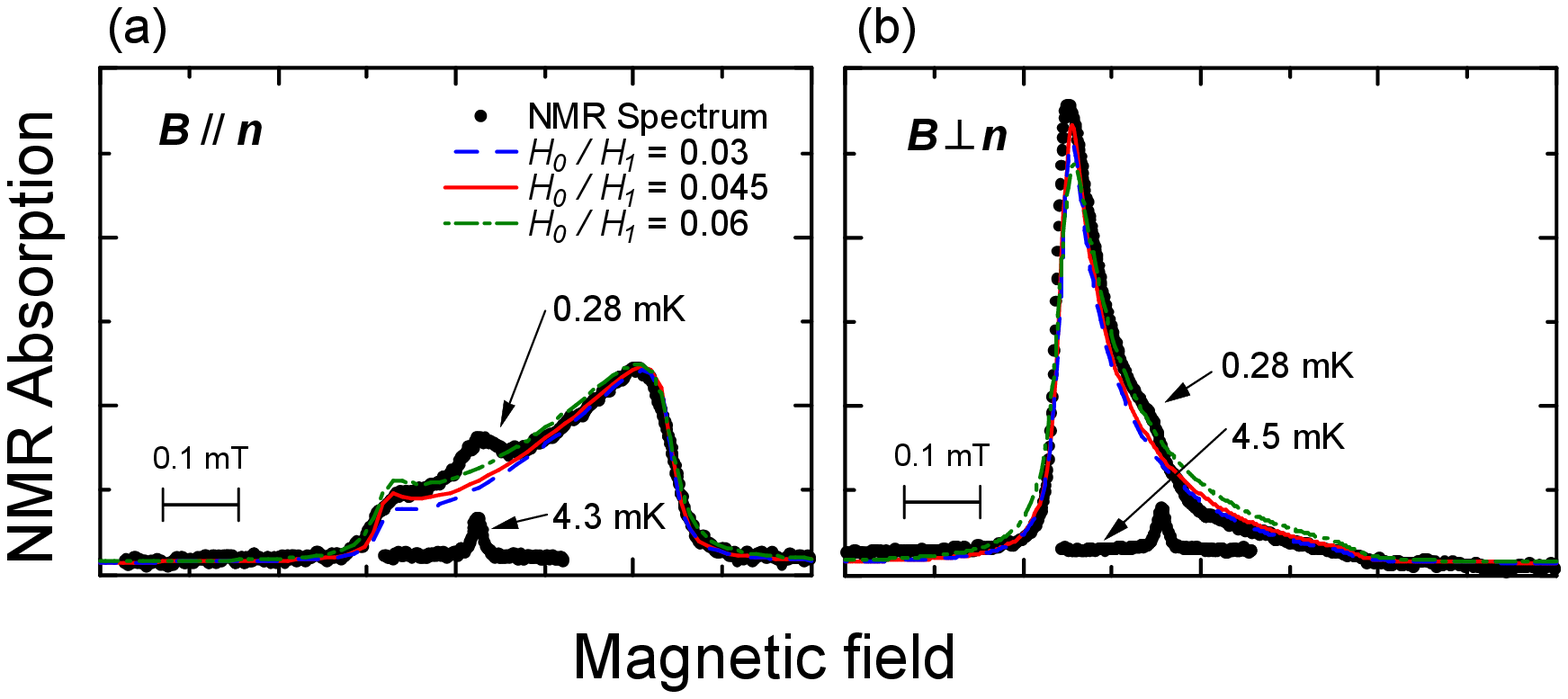}
\end{center}
\caption{(Color online) CW-NMR spectra of the second-layer $^{3}$He adsorbed on Grafoil.
The dashed, solid and dash-dotted lines are the calculated spectra with
$H_{0}/H_{1} = 0.03, 0.045, 0.06$, respectively.
$\delta$ is fixed at 27.5 deg. 
Otherwise, the same as Fig. \ref{fig:cw}.}
\label{fig:cw_param_h}
\end{figure}

\begin{equation}
I(x)\propto
\left\{
\begin{array}{ll}
\displaystyle
\int_{-\pi /2}^{\pi /2}\sin\theta\mathrm{d}\theta\int_{0}^{2\pi}\mathrm{d}\phi
\frac{H(\theta)}{2\pi}
\frac{1}{1+\left\{\frac{x-M(1-3\sin^{2}\theta\sin^{2}\phi)}{A}\right\}^{2}}&
(\boldsymbol{B}\perp\boldsymbol{n})\vspace{3mm}\\
\displaystyle
\int_{-\pi /2}^{\pi /2}\sin\theta\mathrm{d}\theta H(\theta)
\frac{1}{1+\left\{\frac{x-M(1-3\cos^{2}\theta)}{A}\right\}^{2}}&
(\boldsymbol{B}\parallel\boldsymbol{n}).
\end{array}
\right.
\label{eq:cw}
\end{equation}
Here $x \equiv B_{0}- B$, and $A$ is the intrinsic line width.
Thus, $H(\gamma)$ can be determined by fitting the measured NMR spectra
to eq.(\ref{eq:cw}).
We note that similar analyses have been briefly reported
by previous workers \cite{Godfrin_JPhys,Bauerle_PhD}
for the $\boldsymbol{B}\parallel\boldsymbol{n}$ direction.

We fitted the NMR data at $T$ = 0.28 mK for $\rho =$ 21.5 nm$^{-2}$
taken by Schiffer \textit{et al.} \cite{Schiffer_JLTP} 
assuming the gaussian distribution (eq. (\ref{eq:dist_fn})), and obtained 
\begin{equation}
H_{0}/H_{1}=0.045,   \delta =27.5 \;\mathrm{deg}.
\label{eq:dist_nmr}
\end{equation}
The mosaic angle distribution determined by this method is shown in
Fig. \ref{fig:distribution_new}, which is much more anisotropic
than Fig. \ref{fig:distribution}.

The fitted results are shown by the solid lines (red) in Fig. \ref{fig:cw}.
The fitting quality is remarkably good both for (a) $\boldsymbol{B}\parallel\boldsymbol{n}$
and (b) $\boldsymbol{B}\perp\boldsymbol{n}$.
In addition to this, the small fitted value for $H_{0}/H_{1}$ indicates the appropriateness
of the gaussian functional form for $H(\gamma)$. 
Note that there are small bumps near $B_{0}$ in the spectra for both
$\boldsymbol{B}$ directions which are not reproduced by 
the fittings in Fig. \ref{fig:cw}.
These are attributable to magnetization of the paramagnetic first-layer $^{3}$He
\cite{Schiffer_JLTP}.

On the other hand, the dashed lines (blue) in Fig. \ref{fig:cw} are 
spectra calculated from eq. (\ref{eq:dist_neu}), the distribution claimed 
by Taub \textit{et al.} \cite{Taub_PRB}.
They look nearly the same for both directions of $\boldsymbol{B}$ 
because of the large isotropic component $H_{0}$ 
and do not explain the experimental data at all.
We noticed that, in Ref. \cite{Taub_PRB}, they represented the platelet inclination
only with $\theta$ without considering the $\phi$ variation.
We believe that is why their estimation of the mosaic angle distribution
is unrealistically isotropic.

Note that the fitting for the $\boldsymbol{B}\parallel\boldsymbol{n}$ direction
is very sensitive to the presence of the isotropic component $H_{0}$,
while that for the $\boldsymbol{B}\perp\boldsymbol{n}$ is not.
Let us estimate the sensitivity of CW-NMR spectra in determining
the mosaic angle distribution in this method.
In Figs. \ref{fig:cw_param_delta} and \ref{fig:cw_param_h}, we plotted the spectra 
calculated with slightly different $\delta$ and $H_{0}/H_{1}$ values from
those of the best fitting by $\pm 2.5$ deg and $\pm 0.015$, respectively.
The spectra are sensitive enough to discriminate such small variations 
of the distributions.
The constraint that the spectra for both $\boldsymbol{B}\parallel\boldsymbol{n}$
and $\boldsymbol{B}\perp\boldsymbol{n}$ directions should be fitted simultaneously 
with the same parameters enhances the sensitivity.

\section{Conclusions}
\label{sec:conc}
In this paper, we proposed a useful method to estimate the mosaic angle distribution 
of Grafoil platelets from CW-NMR spectra of highly polarized monolayer $^{3}$He 
adsorbed on the substrate.
By analyzing the NMR data taken by the Stanford group \cite{Schiffer_JLTP}, 
we have proved that Grafoil used by them consists of two components.
One is isotropic like powder (about 5 \%) and the other is
distributed in the gaussian function of inclination $\theta$
with a standard deviation $\sigma =27.5\pm2.5$ deg. 
These values are much more anisotropic (or 2D like) than ones determined
from the previous analysis by Taub \textit{et al.} \cite{Taub_PRB}
who used the neutron diffraction data of Ar monolayer adsorbed on Grafoil.
We have found that their integration of the angular distribution
is inappropriate.
The results shown in this paper will be valuable to elucidate
intrinsic 2D properties from various experimental 
data for monolayer systems adsorbed on Grafoil substrate. 
It is desirable to reanalyze their data with the mosaic angle distribution
of Eqs. (\ref{eq:dist_fn}) and (\ref{eq:dist_nmr}) by taking account of
details of the experimental setup.

\begin{acknowledgements}
We thank useful discussions with Bill Mullin. This work was financially 
supported by Grant-in-Aid for Scientific Research on Priority Areas 
(No. 17071002) from MEXT, Japan.
S.T. acknowledges support from the JSPS Research Fellowship.
\end{acknowledgements}

\end{document}